\documentclass[pre,amsfonts,floatfix,superscriptaddress,longbibliography,twocolumn]{revtex4-2}

\usepackage{mathbbol}
\usepackage{amsmath,amssymb,bm,amsthm}
\usepackage{epstopdf}
\usepackage{braket}
\usepackage{soul}

\usepackage[utf8x]{inputenc}
\usepackage[usenames,dvipsnames, pdftex]{xcolor}
\usepackage[colorlinks, linkcolor=blue,citecolor=blue, urlcolor=blue]{hyperref}
\usepackage{graphicx}

\DeclareMathOperator\erfc{erfc}
\DeclareMathOperator\Ai{Ai}

\begin{document}

\title{Ground-state energy distribution of disordered many-body quantum systems} 

\author{Wouter Buijsman} 
\thanks{WB and TLML contributed equally to this work.}
\affiliation{Department of Physics, Ben-Gurion University of the Negev, Beer-Sheva 84105, Israel}

\author{Tal\'ia L. M. Lezama}
\thanks{WB and TLML contributed equally to this work.}
\affiliation{Department of Physics, Yeshiva University, New York, New York 10016, USA}

\author{Tamar Leiser}
\affiliation{Department of Physics, Yeshiva University, New York, New York 10016, USA}

\author{Lea F. Santos}
\affiliation{Department of Physics, Yeshiva University, New York, New York 10016, USA}
\affiliation{Department of Physics, University of Connecticut, Storrs, Connecticut 06269, USA}

\begin{abstract}
Extreme-value distributions are studied in the context of a broad range of problems, from the equilibrium properties of low-temperature disordered systems to the occurrence of natural disasters. Our focus here is on the ground-state energy distribution of disordered many-body quantum systems.
We derive an analytical expression that, upon tuning a parameter, reproduces with high accuracy the ground-state energy distribution of the systems that we consider. 
 For some models, it agrees with the Tracy-Widom distribution obtained from Gaussian random matrices. They include transverse Ising models, the Sachdev-Ye model, and a randomized version of the PXP model. For other systems, such as Bose-Hubbard models with random couplings and the disordered spin-1/2 Heisenberg chain used to investigate many-body localization, the shapes are at odds with the Tracy-Widom distribution. Our analytical expression captures all of these distributions, thus playing a role to the lowest energy level similar to that played by the Brody distribution to the bulk of the spectrum.
\end{abstract}

\maketitle

%%%%%%%%%%%%%%%%%%%%%%%%%%%%%%%%%%%%%%%%%%%%%%%%%%
\section{Introduction}
%%%%%%%%%%%%%%%%%%%%%%%%%%%%%%%%%%%%%%%%%%%%%%%%%%

The level spacing distribution in the bulk of the spectrum of disordered many-body quantum systems and its comparison with full random matrices~\cite{Brody1981,ZelevinskyRep1996,Guhr1998,Kota2001, Buijsman2019} have been intensively investigated, due to their relevance to studies that include the thermalization of isolated quantum systems~\cite{Borgonovi2016,Alessio2016}, many-body localization~\cite{SantosEscobar2004, Nandkishore2015,Alet2018,Abanin2019}, and nonequilibrium quantum dynamics~\cite{Brandino2012,Torres2016Entropy,Cotler2017GUE,Torres2018,Schiulaz2019}. The present work focuses instead on the distribution of the lowest energy level of disordered many-body quantum systems and its relationship with random matrices, a subject that has received less attention~\cite{Gur-Ari2018,Stephan2019}. As our results indicate, agreement with random matrix theory for the energy levels in the bulk of the spectrum does not imply the same for the ground-state energy distribution.

Extreme-value statistics concerns the study of rare events, such as tsunamis, floods, earthquakes, and large variations in the stock market. It has been employed in the context of the Griffiths phase, where rare occurrences of local order take place in an otherwise disordered phase~\cite{Juhasz2006,Kovacs2021}. It involves also the study of the fluctuations of the smallest (largest) eigenvalue of random matrices~\cite{EmbrechtsBook}, which has applications in analyses of the stability of dynamical systems with interactions~\cite{May1972,Majumdar2014} and of the equilibrium properties of disordered systems at low temperatures~\cite{Bouchaud1997,Majumdar2014,Majumdar2020}. In the case of independent and identically distributed random variables, there are three universality classes for the distribution of the sample minimum (maximum). They are given by the Gumbel, Fr\'echet, or Weibull distribution, depending on the tail of the parent probability density of the variables~\cite{Majumdar2020,Leadbetter1983}. When, instead, the random variables are correlated, there are few cases for which the extreme-value distribution has been obtained~\cite{Majumdar2020,Leadbetter1983}, one being the distribution derived by Tracy and Widom~\cite{Tracy1994,Tracy1996} for the lowest (highest) eigenvalue of ensembles of large Gaussian random matrices.

The Tracy-Widom distribution arises in different theoretical and experimental contexts, such as in studies of mesoscopic fluctuations in quantum dots and of spatial correlations of noninteracting fermions at the edges of a trap (see Refs.~\cite{Bouchaud1997, Majumdar2020} and references therein). Of particular interest to us is the verification that the ground-state energy distribution of even-even nuclei~\cite{Johnson1998,Kusnezov2000,Zhao2004} agrees with the Tracy-Widom distribution obtained with Gaussian orthogonal ensembles (GOEs)~\cite{Santos2002}.
Nuclei are typical examples of interacting many-body quantum systems, so features found there may extend to other similar models. This prompts us to investigate the distribution of the lowest level of different disordered many-body quantum systems. 

We find that the ground-state energy distribution of various spin models with short- and long-range random couplings, such as transverse Ising models~\cite{Lin2017}, the Sachdev-Ye model~\cite{Sachdev1993}, and a randomized version of the PXP model~\cite{Lesanovsky2012} is comparable to the Tracy-Widom distribution. Many of these systems can be realized in experiments with cold atoms~\cite{Bernien2017,Bloch2008,Kaufman2016}, ion traps~\cite{Blatt2012,Richerme2014}, and nuclear magnetic resonance~\cite{Wei2018,Niknam2020}.
However, we also identify examples of disordered many-body quantum systems, where the ground-state energy distribution is different from the Tracy-Widom distribution. These include spin-1/2 Heisenberg chains with onsite disorder and Bose-Hubbard models with random couplings. 

The disagreement with the Tracy-Widom distribution can happen even when the level spacing distribution in the middle of the spectrum concurs with random matrix theory. Indeed, for the disordered spin-1/2 Heisenberg chain, as the disorder strength increases and the level spacing distribution for the bulk of the spectrum moves from the Wigner-Dyson distribution of random matrix theory to the Poissonian shape of integrable models~\cite{Santos2004}, we see that the ground-state energy distribution changes in the opposite direction, from a shape far from the Tracy-Widom distribution to a form very similar to it.

We derive a general one-parameter analytical expression that reproduces up to high accuracy the ground-state energy distribution of all the random models that we study, except for the Bose-Hubbard model with a {\em fixed} tunneling strength. Similar to the distributions proposed by Brody {\it et al.}~\cite{Brody1981} and Izrailev~\cite{Izrailev1990} that interpolate between the Poisson and the Wigner-Dyson distribution for the bulk of the spectrum, our expression captures all the shapes of the ground-state energy distribution reached by the disordered spin-1/2 Heisenberg chain as the disorder strength is changed. 

The outline of this work is as follows. Section~\ref{sec: RMT} presents the extreme value statistics for large and $2 \times 2$ Gaussian random matrices. In Sec.~\ref{sec: agreement}, we compare the ground-state energy distribution of various models with the random matrices' results. In Sec.~\ref{sec: expression}, we introduce our single-parameter expression for the distributions of ground-state energies. In Sec.~\ref{sec: chi2}, the agreement between our expression and the ground-state energy distribution of physical models as well as of random matrices is illustrated numerically. Conclusions are given in Sec.~\ref{sec: conclusions}.

%%%%%%%%%%%%%%%%%%%%%%%%%%%%%%%%%%%%%%%%%%%%%%%%%%
\section{Ground-state energy distributions from random matrices} \label{sec: RMT}
%%%%%%%%%%%%%%%%%%%%%%%%%%%%%%%%%%%%%%%%%%%%%%%%%%
We first review the ground-state energy distribution of $N \times N$ GOE random matrices for $N \to \infty$, which is the Tracy-Widom distribution, and for $N=2$, which has been compared with the ground-state energy distribution of some nuclear and molecular models~\cite{Santos2002}.

\subsection{Tracy-Widom distribution}
The GOE consists of real-valued symmetric matrices with entries sampled independently from the Gaussian distribution with zero mean $\mu = 0$ and off-diagonal (diagonal) components with variance $\sigma^2 = 1/2$ ($\sigma^{2} = 1$) \cite{MehtaBook,Livan2018}. The eigenvalues $E_0, E_1, \dots, E_{N-1}$ are distributed according to
\begin{equation}
P(E_0, E_1, \dots, E_{N-1}) = \frac{1}{\mathcal{Z}_N} \prod_{i=0}^{N-1} e^{- E_i^2 / 2} \prod_{j < k} | E_j - E_k |,
\label{eq: P-GOE}
\end{equation}
where $\mathcal{Z}_N$ is a normalization constant that fixes the probability integral to unity. This expression does not factorize in terms that uniquely depend on a single eigenvalue, indicating that the eigenvalues are (strongly) correlated. Sorting the eigenvalues in increasing order, the distribution of the smallest eigenvalue $E_0$ is obtained by integrating out all other eigenvalues,
\begin{equation}
P(E_0) = \int_{-\infty}^\infty P(E_0, E_1,\dots,E_{N-1}) \, dE_1 \, dE_2 \dots dE_{N-1}.
\end{equation}
For $N \to \infty$, the distribution of the smallest eigenvalue converges towards the Tracy-Widom distribution \cite{Tracy1994, Tracy1996}.
This distribution does not have a closed-form expression, but it can be written in terms of the solution of the Painlev\'e II differential equation and evaluated numerically. After some manipulation~\cite{Finch2018}, it is found that the distribution of the ground-state energy has the shape
\begin{equation}
P(E_0) = \sqrt{F_2(-E_0)} \, \exp \bigg[ \frac{1}{2} \int_{-E_0}^\infty q(x) dx \bigg]
\label{Eq:TW}
\end{equation}
with
\begin{equation}
F_2(x) = \exp \bigg[ - \int_{x}^\infty (z - x) q^2(x) dz \bigg],
\end{equation}
where $q(x)$ is the solution of the Painlev\'e II differential equation $q'' = x q + 2 q^3$ subjected to the boundary condition $q(x) \sim \Ai(x)$ for $x \to \infty$, with $\Ai(x)$ denoting the Airy function.

\subsection{$\pmb{N=2}$ GOE matrices}

The analytical expression of $P(E_0)$ for the $2\times2$ GOE random matrices can be directly derived from Eq.~\eqref{eq: P-GOE} with $N=2$ and it is given by~\cite{Santos2002}
\begin{equation}
P(E_0) = \frac{1}{2 \sqrt{\pi}} e^{-E_0^2} - \frac{1}{2 \sqrt{2}} e^{- \frac{1}{2} E_0^2} E_0 \erfc \bigg(\frac{E_0}{\sqrt{2}} \bigg),
\label{eq:2x2}
\end{equation}
where $\erfc(x)$ is the complementary error function.

%==============Figure 1============
\begin{figure*}[htb!]
\centering
\includegraphics[width=1\textwidth]{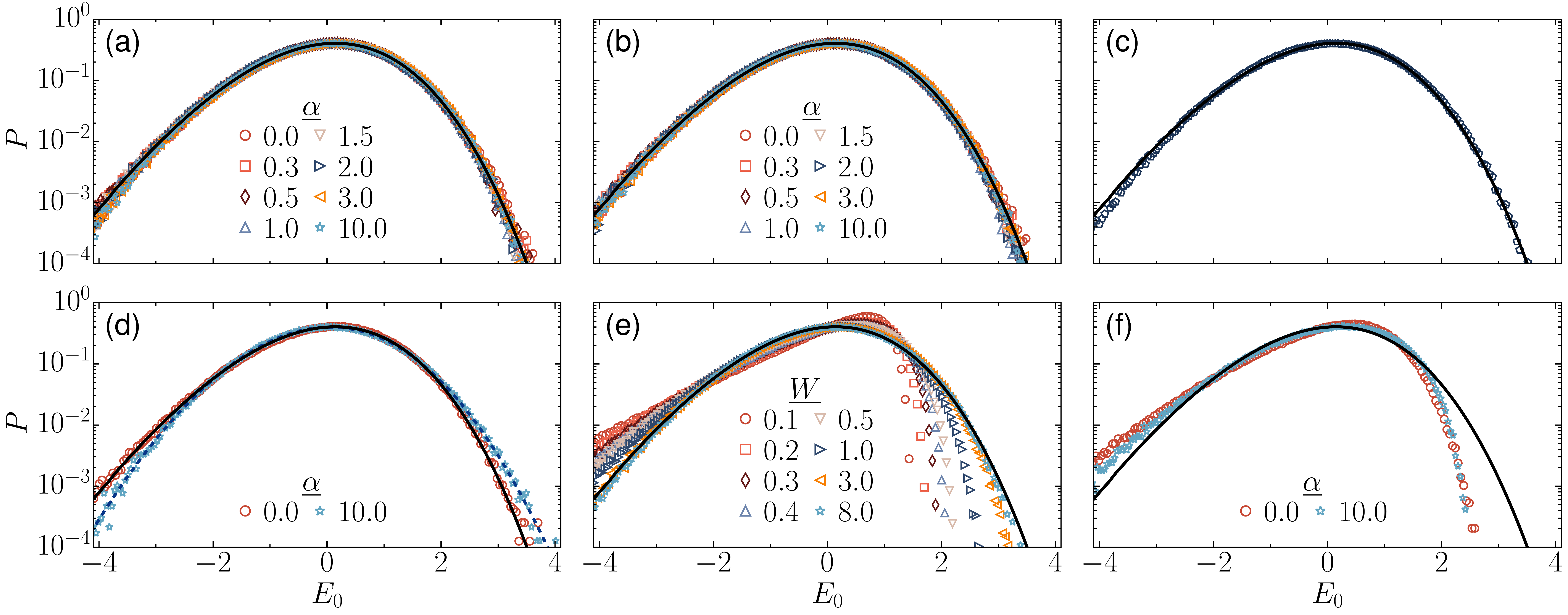}
\caption{Ground-state energy distribution (a)-(f) of the systems defined in Eqs.~(\ref{Eq:SYK})-(\ref{Eq:BH}). The data (symbols) consist of $10^{6}$ realizations for (a) $H$ in Eq.~(\ref{Eq:SYK}) for the zero-magnetization sector (${\cal S}_z=0$) and the system size $L=16$, (b) $H$ in Eq.~(\ref{Eq:Ising}) for $L=14$, (c) $H$ in Eq.~(\ref{Eq:PXP}) for $L=16$, (d) as in panel (a) but with random parameters drawn from a box distribution, (e) $H$ in Eq.~(\ref{Eq:MBL}) for ${\cal S}_z=0$ and $L=16$, and (f) $H$ (\ref{Eq:BH}) with $N_b=L=7$. The solid lines represent the Tracy-Widom distribution and the dashed lines in panel (d) represent the $\mathrm{GOE}_{2\times 2}$ ground-state energy distribution.}
\label{Fig:1}
\end{figure*}
%=================================

%%%%%%%%%%%%%%%%%%%%%%%%%%%%%%%%%%%%%%%%%%%%%%%%%%
\section{Physical systems vs random matrix results} 
\label{sec: agreement}
%%%%%%%%%%%%%%%%%%%%%%%%%%%%%%%%%%%%%%%%%%%%%%%%%%
We compare the Tracy-Widom distribution in Eq.~(\ref{Eq:TW}) with the ground-state energy distribution $P(E_{0})$ of different disordered many-body quantum systems. For all the models considered in this work, the random parameters are independent real numbers sampled from the Gaussian distribution with mean $\mu = 0$ and variance $\sigma^2 = 1$, as in GOE matrices, except when stated otherwise. 

When studying extreme value statistics, it is convenient to shift the distribution such that the mean equals zero, and to scale the distribution such that the variance is unity \cite{Leadbetter1983, Bouchaud1997, Majumdar2020}. Shifting and scaling the distributions such that they have a fixed first (mean) and second (variance) moment eliminates the effects due to characteristic energy scales, allowing one to compare different types of systems on an equal footing. This is what we do in all of our figures below.

In Figs.~\ref{Fig:1}(a)-\ref{Fig:1}(c), we show examples of models whose ground-state energy distributions are very close to the Tracy-Widom distribution, and in Figs.~\ref{Fig:1}(d)-\ref{Fig:1}(f), we show examples of deviations. 

In Fig.~\ref{Fig:1}(a), we plot $P(E_0)$ for the Hamiltonian
\begin{equation}
H = \sum_{i=1}^L \frac{\epsilon_i}{2} \sigma_i^z + \sum_{i,j}^{L-1} \frac{J_{ij}}{4|j-i|^{\alpha}} \vec{\sigma}_i \cdot \vec{\sigma}_j ,
\label{Eq:SYK}
\end{equation}
where $L$ is the system size, $\epsilon_i$ and $J_{ij}=J_{ji}$ are random numbers, and $\sigma^{x,y,z}$ are the Pauli matrices. For all-to-all couplings ($\alpha=0$), the Hamiltonian in Eq.~(\ref{Eq:SYK}) is analogous to the two-body random ensembles studied in quantum chaos and nuclear physics~\cite{French1970,Bohigas1971,Brody1981,Kota2001,Papenbrock2006} and coincides with the Sachdev-Ye model when $\epsilon_i=0$ \cite{Sachdev1993}. We also study $\alpha>0$, in which case we consider one-dimensional (1D) systems. Hamiltonian~(\ref{Eq:SYK}) conserves the total $z$-magnetization ${\cal S}_z = \sum_{i=1}^L \sigma_i^z / 2$. 

As seen in Fig.~\ref{Fig:1}(a), the distribution of the ground-state energy of $H$ (\ref{Eq:SYK}) is very close to the Tracy-Widom distribution, especially for $\alpha=0$, and as $\alpha$ increases from all-to-all to short-range couplings, the deviations remain minor. For the ${\cal S}_z =0$ sector, the agreement holds for system sizes as small as $L=8$; and for large system sizes, there is very good agreement for as few as four excitations. $P(E_0)$ is also close to the Tracy-Widom distribution for the Sachdev-Ye model, that is, for $H$ (\ref{Eq:SYK}) with $\alpha=0$ and $\epsilon_i=0$.

The good agreement with the Tracy-Widom distribution holds also in Figs.~\ref{Fig:1}(b) and \ref{Fig:1}(c).
In Fig.~\ref{Fig:1}(b), the comparison is made with the Ising model with longitudinal and transverse fields,
\begin{equation}
H = \sum_{i=1}^L \left( \frac{h^z_i}{2} \sigma_i^z + \frac{h^x_i}{2} \sigma_i^x \right) + \sum_{i,j}^{L-1} \frac{J_{ij}}{4|j-i|^{\alpha}} \sigma_i^z \sigma_j^z ,
\label{Eq:Ising}
\end{equation}
where $h^z_i$, $h^x_i$, and $J_{ij}=J_{ji}$ are random numbers. In the case of nearest-neighbor couplings, Eq.~(\ref{Eq:Ising}) agrees with the Hamiltonian used to study quantum phase transitions in Ref.~\cite{Lin2017}. We examine both short- and long-range interactions.
In Fig.~\ref{Fig:1}(c), we consider the Hamiltonian
\begin{equation}
H = \sum_{i=1}^{L-2} J_{i+1} P_i \sigma^x_{i+1} P_{i+2} + J_1 \sigma^x_1 P_2 + J_L P_{L-1} \sigma^x_L ,
\label{Eq:PXP}
\end{equation}
where $P_i = (1 - \sigma^z_i)/2$ denotes the projection operator and $J_i$ are random numbers. The main mechanism of this model is based on local constraints that forbid two adjacent spins to be simultaneously in the up-state. For $J_i = 1$, the Hamiltonian reduces to the PXP model \cite{Lesanovsky2012,Bernien2017}, which is paradigmatic in studies on quantum many-body scars and weak ergodicity breaking~\cite{Turner2018,Turner2018b}. For random parameters, model (\ref{Eq:PXP}) is similar to the one introduced in Ref.~\cite{Chen2018}.

In Fig.~\ref{Fig:1}(d), we analyze again the ground-state energy distribution of $H$~\eqref{Eq:SYK}, but now with random parameters $\epsilon_{i}$ and $J_{ij}$ drawn from a box distribution. The good agreement with the Tracy-Widom distribution persists for all-to-all couplings ($\alpha=0$), although by decreasing the range of the interactions ($\alpha \rightarrow \infty $), the distribution approaches the one obtained with the $2 \times 2$ random matrices [Eq.~(\ref{eq:2x2})]. This sensitivity to the choice of random numbers contrasts with what one finds at the bulk of the spectrum and deserves further investigation~\cite{footNote}. Below and everywhere else in this work, we consider only random numbers from Gaussian distributions.

All the models considered up to this point are different from GOE random matrices. They involve only two-body couplings, their Hamiltonian matrices are sparse, and their elements, despite being random, present correlations, many of them being identical. Yet, we have found ground-state energy distributions similar to those obtained with large or $2\times 2$ random matrices. In what follows, we fix the coupling parameter $J$ in Eq.~(\ref{Eq:SYK}) and consider only nearest-neighbor couplings, thus moving even further from GOE random matrices.

In Fig.~\ref{Fig:1}~(e), we analyze the disordered spin-1/2 Heisenberg Hamiltonian employed in studies of many-body localization~\cite{SantosEscobar2004,Nandkishore2015,Alet2018,Abanin2019}, 
\begin{equation}
H = W\sum_{i=1}^L \frac{ h_i}{2} \sigma_i^z + \frac{J}{4} \sum_{i=1}^{L-1} \vec{\sigma}_i \cdot \vec{\sigma}_{i+1}.
\label{Eq:MBL}
\end{equation}
In the equation above, $h_i$ are random numbers, $W$ is the disorder strength, and the coupling parameter $J=1$. Interestingly, even this model, which has a very sparse Hamiltonian matrix and a high level of correlated matrix elements, still presents ground-state energy distributions similar to the Tracy-Widom distribution when the disorder strength is large ($W>J$). However, as the disorder strength decreases ($0< W \leq J$), the lowest-energy distributions get more skewed and different from the random matrices results. This is in stark contrast with the properties of the bulk of the spectrum of this model, which are similar to those of GOE random matrices when $0<W \leq J$, but differ when $W>J$ \cite{Avishai2002,Santos2004}.

In Secs.~\ref{sec: expression} and \ref{sec: chi2} below, we resort to real-valued Wishart ensembles and identify the distributions that best match the various shapes shown in Fig.~\ref{Fig:1}~(e). Other models with a fixed constant parameter, such as the Hamiltonians in Refs.~\cite{Chen2018,Rademaker2020}, also deviate from the Tracy-Widom distribution, but display shapes that can be captured with the expression presented in Sec.~\ref{sec: expression}. The only model with a fixed parameter that we studied here that does not follow this rule is the disordered 1D Bose-Hubbard model discussed next.

The Hamiltonian of the disordered 1D Bose-Hubbard model is given by
\begin{equation}
H = \sum_{i=1}^{L} \frac{U_{i} }{2 } n_i\left(n_i-1\right) - \sum_{i,j}^{L-1}
\frac{J_{i,j}}{|j-i|^{\alpha}} \left(a_i^\dagger a_j + a_j^\dagger a_i\right),
\label{Eq:BH}
\end{equation}
where $a_i$ $(a_i^\dagger)$ is the annihilation (creation) operator, $n_i = a_i^\dagger a_i$, 
and $U_{i}$ are random numbers. 
For deterministic tunneling, $J_{i,j}=1$, the ground-state energy distribution displays a kink that becomes more pronounced as $\alpha \rightarrow 0$ (see Appendix~A) and which cannot be reproduced with the analytical expression in Sec.~\ref{sec: expression}. This behavior changes, however, when both $U_{i}$ and  $J_{i,j}$ are random numbers. In this case, when the total number of bosons $N_b=L$, the distribution deviates from the Tracy-Widom distribution for both short- and long-range couplings, as seen in Fig.~\ref{Fig:1}(f), but $P(E_0)$ is well captured with the expression in Sec.~\ref{sec: expression}. In addition, when  $N_b$ is small, $P(E_0)$ shows good agreement with the Tracy-Widom distribution.

Overall, our results below suggest that the single-parameter expression in Sec.~\ref{sec: expression} reproduces well the ground-state energy distributions of most many-body quantum systems considered here. It shows good agreement with the Tracy-Widom distribution, and by tuning the parameter, the expression captures also the distributions that cannot be matched by the Tracy-Widom distribution.

%%%%%%%%%%%%%%%%%%%%%%%%%%%%%%%%%%%%%%%%%%%%%%%%%%
\section{Analytical expression: \pmb{$\chi^2$} distribution} 
\label{sec: expression}
%%%%%%%%%%%%%%%%%%%%%%%%%%%%%%%%%%%%%%%%%%%%%%%%%% 
The Wishart ensemble is one of the classical ensembles in random matrix theory and precedes the Gaussian ensembles \cite{Livan2018}. It consists of matrices $A = X X^T$, where $X$ is an $N \times M$ matrix with (real-valued) entries sampled independently from the Gaussian distribution with mean $\mu = 0$ and variance $\sigma^2 = 1$. The eigenvalues $E_0, E_1, \dots, E_{N-1}$ are distributed according to
\begin{equation}
\begin{split}
 & P(E_0, E_1, \dots, E_{N-1}) = \frac{1}{\mathcal{Z}_N} \prod_{i=0}^{N-1} e^{- E_i/2}  \\
 & \times \prod_{i=0}^{N-1} E_i^{(M - N - 1)/2}
\prod_{j < k} | E_j - E_k|, \quad (E_i \ge 0),
\label{eq: P-Wishart}
\end{split}
\end{equation}
where, as in Eq.~\eqref{eq: P-GOE}, the prefactor $1 / \mathcal{Z}_N$ is a normalization constant fixing the integral probability to unity. 

For $N=1$, the distribution of the single, therefore extreme value, is given by
\begin{equation}
P(E_0) = \frac{1}{2^{M/2} \Gamma(M/2)} e^{-E_0 / 2} E_0^{M/2 - 1}, \quad \,\, (E_0 \ge 0).
\label{eq: P-Wishart-N1}
\end{equation}
This equation coincides with the $\chi^2$-distribution with $M$ degrees of freedom, that is, it corresponds to the distribution of the sum of $M$ samples from the standard Gaussian distribution. $P(E_0)$ in Eq.~(\ref{eq: P-Wishart-N1}) has mean $\mu = M$ and variance $\sigma^2 = 2M$, and it remains well-defined even when the free parameter $M$ is noninteger. 

%==============Figure 2============
\begin{figure}[t!]
\centering
\includegraphics[width=1\columnwidth]{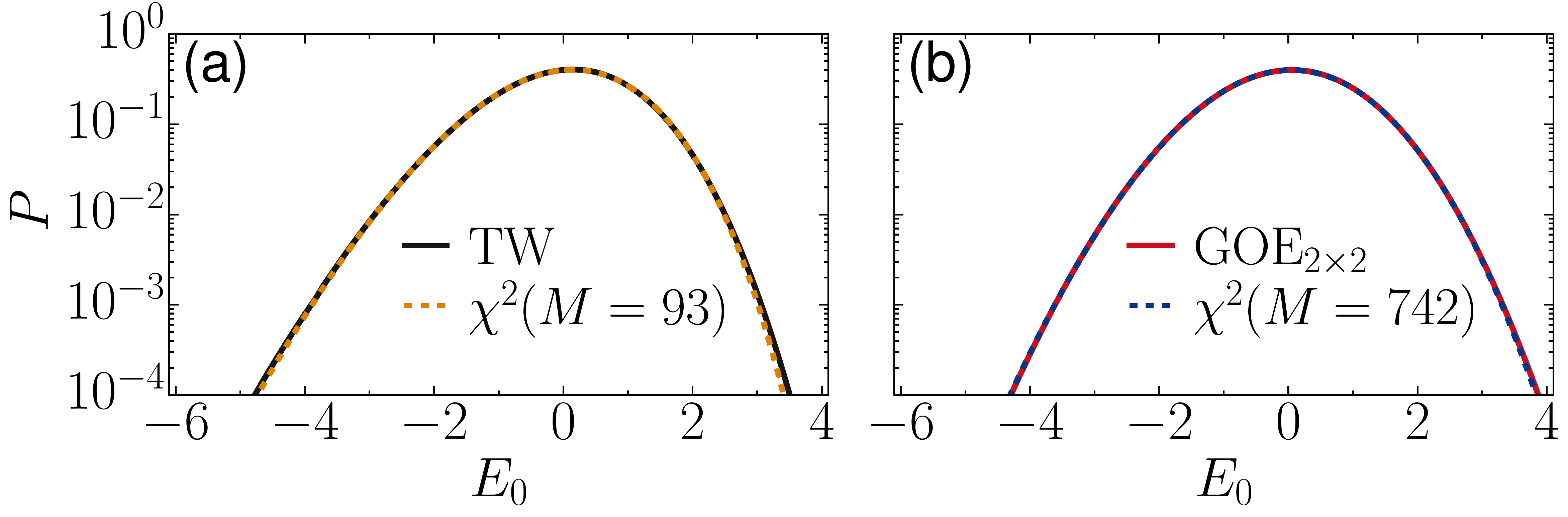}
\caption{Comparison of the Tracy-Widom distribution  (a) and the $2\times 2$ GOE ground-state energy distribution  (b) with the best-fitting $\chi^2$ distribution. The fitting values are $M \approx 92.89$ and $M \approx 741.60$, respectively (these values are rounded to their closest integers when written in the panels).
}
\label{Fig:2}
\end{figure}
%=================================

%==============Figure 3============
\begin{figure*}[htb!]
\centering
\includegraphics[width=1\textwidth]{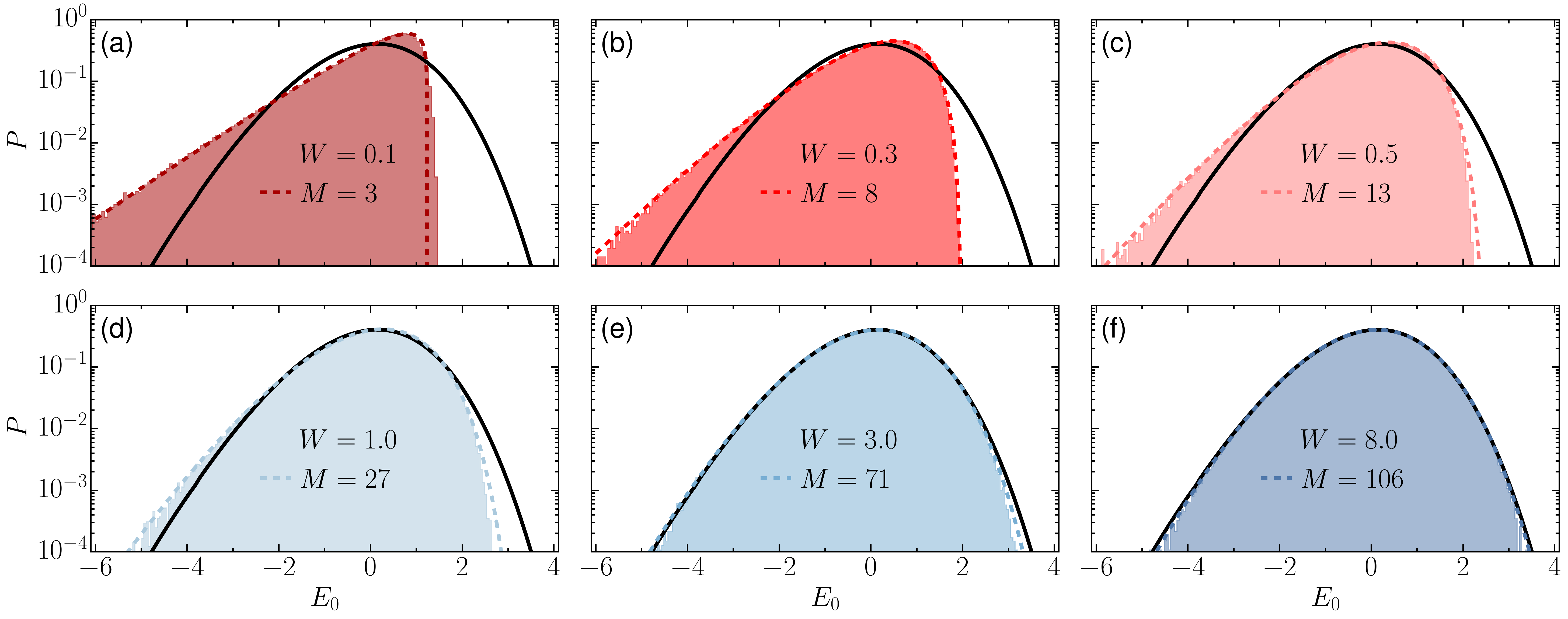}
\caption{Ground-state energy distribution of the Hamiltonian in Eq.~(\ref{Eq:MBL}), which is prototypical in studies of many-body localization. The distribution deviates from (approaches) the Tracy-Widom distribution for disorder strength $0<W \leq J$ ($W > J$). The data (shades) are obtained using $10^{6}$ realizations for the zero-magnetization sector, ${\cal S}_z=0$, and system size $L=16$. Solid and dashed lines represent the Tracy-Widom distribution and the $\chi^2$ distribution, respectively. The fitting values of $M$ corresponding to each panel from (a) to (f) are: $M \approx 3.04, 7.80, 12.93, 26.73, 71.42,$ and $106.27$ (these values are rounded to their closest integers when written in the panels).
}
\label{Fig:3}
\end{figure*}
%=================================

Our focus is on the minimum, instead of the maximum, so in Eq. \eqref{eq: P-Wishart-N1} we consider the distribution of $-E_0$. Since the first (mean) and second (variance) moments of all distributions are fixed by shiftings and scalings, the lowest moment that is not fixed is the third one (skewness). To find the best-fitting value of $M$ for a given ground-state energy distribution, we use the skewness $\gamma = -\mu_3 / \mu_2^{3/2}$ 
as a matching parameter, taking into account that the $n$-th moment is $\mu_n = \int (x - \mu)^n P(x) \, dx$. We then have that $\gamma = -2 \sqrt{2 /M}$ and thus
\begin{equation}
M = \frac{8}{\gamma^2}.
\label{eq:skewness-M}
\end{equation}
By varying $M$, the $\chi^2$-distribution can be tuned to any skewness and, as we show below, it can then match any of the ground-state energy distributions obtained in this work,  except for the 1D Bose-Hubbard model with a fixed tunneling parameter.

%%%%%%%%%%%%%%%%%%%%%%%%%%%%%%%%%%%%%%%%%%%%%%%%%%
\section{Agreement with the $\pmb{\chi^2}$ distribution} 
\label{sec: chi2}
%%%%%%%%%%%%%%%%%%%%%%%%%%%%%%%%%%%%%%%%%%%%%%%%%%
We start by comparing the $\chi^2$ distribution in Eq.~(\ref{eq: P-Wishart-N1}) with the Tracy-Widom distribution. The skewness of the Tracy Widom distribution has been obtained numerically as $\gamma \approx -0.293$ \cite{Finch2018,noteFinch}. Equating this value to the skewness of the $\chi^2$ distribution gives $M \approx 92.89$.  As shown in Fig.~\ref{Fig:2}(a), the two curves agree down to very small values. The parallel between these two distributions is useful, because a closed form for the Tracy-Widom distribution does not yet exist. One needs to evaluate it numerically, which requires solving the Painlev\'e II differential equation, so the $\chi^2$ distribution is a simple and accurate alternative (see also Ref.~\cite{Chiani2014}).

By further increasing $M$ beyond the Tracy-Widom shape, the $\chi^2$ distribution eventually matches the analytical expression of $P(E_0)$ obtained with $2\times2$ GOE random matrices and given in Eq.~\eqref{eq:2x2}. The skewness of this distribution,
$\gamma = - [2(\pi - 3) \sqrt{\pi}/(6-\pi)^{3/2}]
$,
matches the skewness of the $\chi^2$-distribution for $M \approx 741.60$. The excellent agreement between the two curves is illustrated in Fig.~\ref{Fig:2}~(b). Finally, for $M \to \infty$, the $\chi^2$ distribution converges towards a Gaussian.

In Fig.~\ref{Fig:3}, we select some of the curves exhibited in Fig.~\ref{Fig:1}(e) for the spin-1/2 model used to study many-body localization and show that the analytical expression in Eq.~(\ref{eq: P-Wishart-N1}) matches the numerical results for $P(E_0)$ obtained with the Hamiltonian in Eq.~(\ref{Eq:MBL}) for any disorder strength $W$. The panels span a broad range of fitted values of $M$, thus forming a representative selection. For $W=0.1$ in Fig.~\ref{Fig:3}(a), the shape of the distribution is highly skewed and similar to what one finds for the Weibull distribution~\cite{PapoulisBook}, while for $W=8$ in Fig.~\ref{Fig:3}(f), the numerical result gets close to the Tracy-Widom distribution.  The quantification of the agreement between the numerical results and the fitted $\chi^2$ distribution is done in Appendix \ref{App1} through the difference between the compared curves and through their higher-order moments. 

%==============Figure 4============
\begin{figure}[h!]
\centering
\includegraphics[width=1.0\columnwidth]{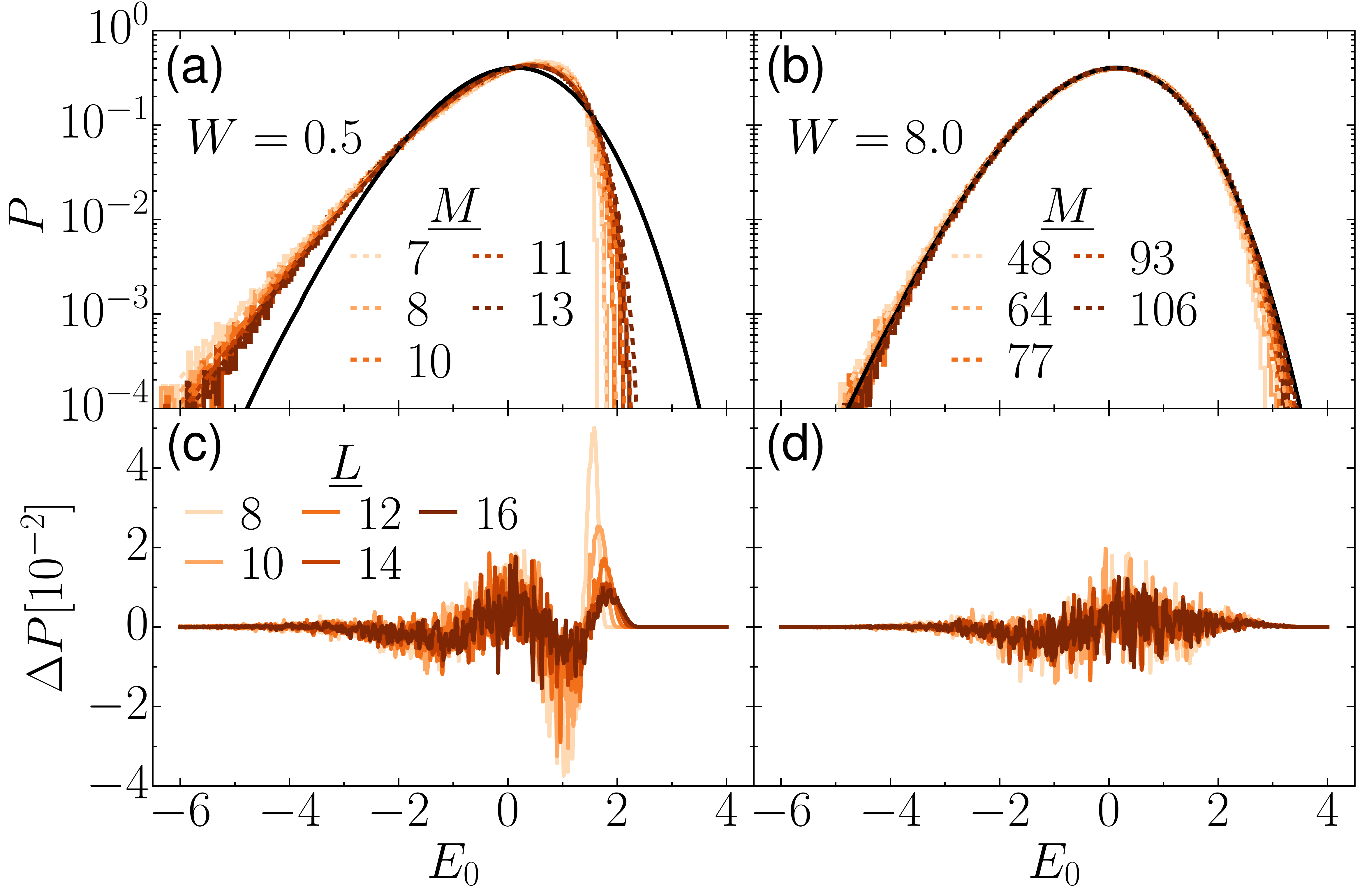}
\caption{Finite-size scaling of the ground-state energy distribution of $H$ \eqref{Eq:MBL}. For each system size, $P(E_0)$ is obtained with $10^{6}$ disorder realizations for disorder strengths (a) $W = 0.5$ and (b) $W=8.0$. Solid and dashed lines represent the Tracy-Widom and the best-fitting $\chi^{2}$ distributions, respectively. The fitting values of $M$ corresponding to each $L$ in increasing order are as follows: $M \approx 6.51, \, 8.10, \, 9.74, \, 11.28,$ and $12.93$ for $W=0.5$, and $M \approx 47.76, \, 77.01, \, 93.32,$ and $106.27$ for $W=8.0$ (these values are rounded to their closest integers when written in the panels). Panels (c) and (d) show a difference of the order of~$10^{-2}$ between the distributions and their best fitting $\chi^{2}$-distributions, $\Delta P(E_{0}) \equiv P(E_{0})-P_{\chi^{2}}(E_{0})$, for the data in Figs.~\ref{Fig:4}(a) and \ref{Fig:4}(b), respectively.
}
\label{Fig:4}
\end{figure}
%=================================

In Fig.~\ref{Fig:4}, we examine the ground-state energy distribution of the system described in Eq.~\eqref{Eq:MBL} as a function of system size. We observe that, as the system size becomes larger, the value of $M$ for the best fitting $\chi^{2}$ distribution increases, as seen for $W=0.5$ in Fig.~\ref{Fig:4}(a) and $W=8$ in Fig.~\ref{Fig:4}(b). Our results also suggest that the agreement between the system's ground-state energy distribution and the best-fitting distribution improves with increasing system size, as quantified by their difference $\Delta P(E_{0}) \equiv P(E_{0})-P_{\chi^{2}}(E_{0})$ in Figs.~\ref{Fig:4}(c) and \ref{Fig:4}(d). For $W=8$ [Fig.~\ref{Fig:4}(d)], in particular, we observe an overall faster convergence and little difference between system sizes $L=14$ and $L=16$.
However, with the available numerical data, it is not possible to state whether there will be convergence to a specific shape in the thermodynamic limit and how this shape might depend on the model and its parameters. 

The problem of convergence in the thermodynamic limit remains an open question also in the analysis of level spacing distributions in the bulk of the spectrum, where scaled-up numerical studies have suggested that an infinitesimal integrability breaking term may lead to the Wigner-Dyson distribution~\cite{Santos2010PRE,Santos2020}, but no proof exists yet. For finite systems, however, just as the Brody and the Izrailev distributions reproduce any of the level spacing distributions between Poisson and Wigner-Dyson, the $\chi^{2}$ distribution captures any of the lowest-energy distributions shown in Figs.~\ref{Fig:1}-\ref{Fig:3}, from a highly skewed shape, as in Fig.~\ref{Fig:3}(a), to the random matrices results in Figs.~\ref{Fig:2}(a) and \ref{Fig:2}(b), and up to the Gaussian shape of a diagonal random matrix.

%%%%%%%%%%%%%%%%%%%%%%%%%%%%%%%%%%%%%%%%%%%%%%%%%%
\section{Conclusions} 
\label{sec: conclusions}
%%%%%%%%%%%%%%%%%%%%%%%%%%%%%%%%%%%%%%%%%%%%%%%%%%
Motivated by studies in nuclear physics, we compared the ground-state energy distribution of random matrices (Tracy-Widom distribution) with several physical models of current experimental interest. There was no indication of a correspondence between the results for the bulk of the spectrum with those for the lowest level. In the case of the Heisenberg spin-1/2 chain with onsite disorder, for example, when the level statistics in the bulk of the spectrum agrees with random matrix theory, the lowest-energy distribution disagrees, and vice-versa.

After realizing that the Tracy-Widom distribution does not always match the ground-state energy distribution of disordered many-body quantum systems, we searched for an alternative that could capture those cases and could also show good agreement with the Tracy-Widom distribution, thus providing a general picture.

We found that by appropriately tuning the degree of skewness of  the $\chi^2$ distribution, it reproduces well the ground-state energy distributions of all of the models with random couplings that we considered. This was also the case for the disordered Heisenberg spin-1/2 chain with a fixed coupling strength. 
 
Finding the best fitting value of the skewness that ensures agreement between the $\chi^2$ distribution and the lowest-energy distribution of finite disordered many-body quantum systems plays a role similar to fitting the Brody distribution to match the level spacing distribution of the bulk of the spectra of those systems. An open question shared by these two types of analyses is the convergence of the distributions in the thermodynamic limit.

%%%%%%%%%%%%%%%%%%%%%%%%%%%%%%%%%%%%%%%%%%%%%%%%%%%%%%
\begin{acknowledgments}
This research was supported by the United States National Science Foundation (NSF) Grant No. DMR-1936006. WB acknowledges support from the Kreitman School of Advanced Graduate Studies at Ben-Gurion University. L.F.S. had support from the MPS Simons Foundation Award ID: 678586.

\end{acknowledgments}

\vspace{2cm}

%%%%%%%%%%%%%%%%%%%%%%%%%%%%%%%%%%%%%%%%%%%%%%%%%%%%%%
\appendix
\section{Bose-Hubbard Model} 
\label{AppBH}

\begin{figure}[t!]
\centering
\includegraphics[width=1.0\columnwidth]{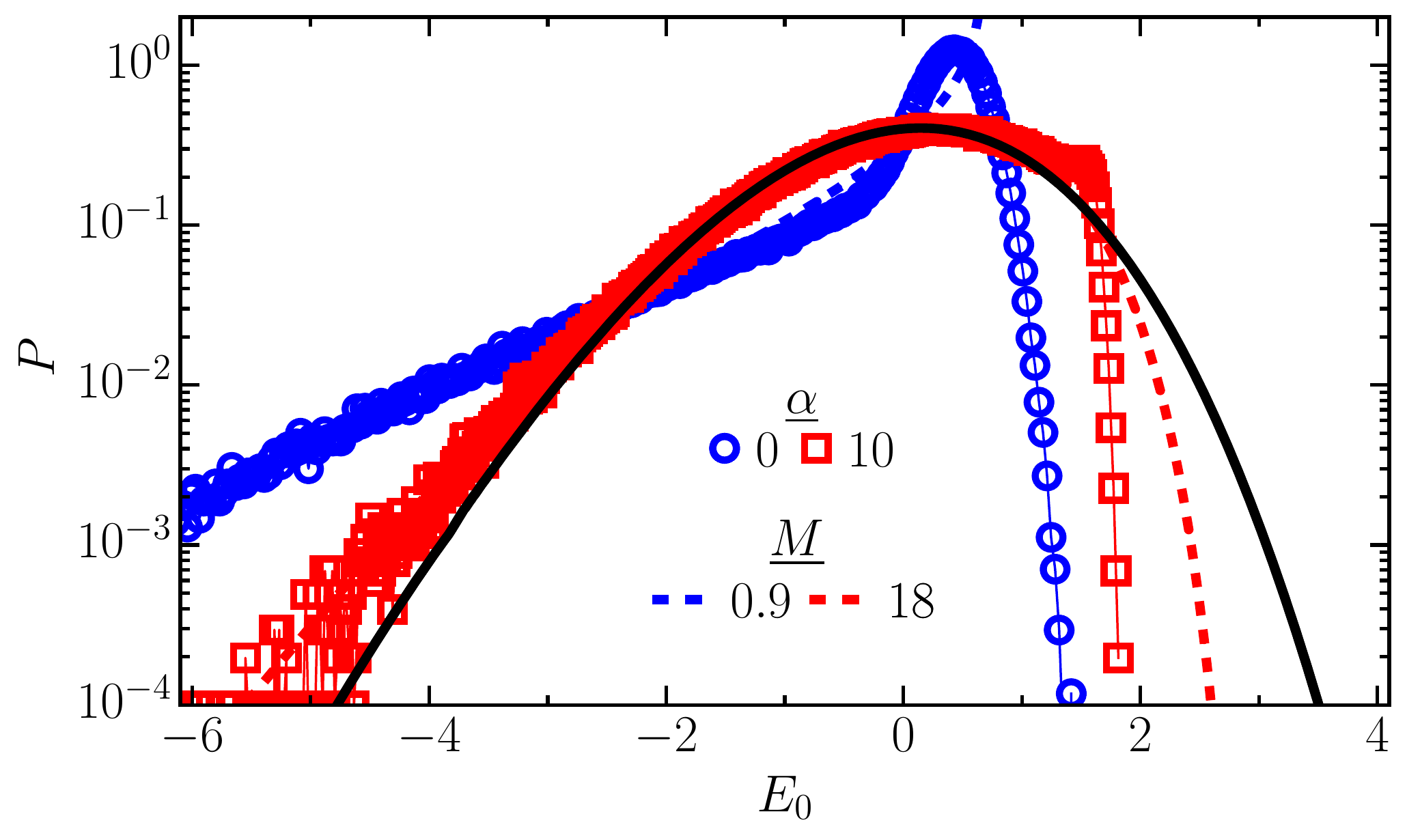}
\caption{Ground-state energy distribution for the Bose-Hubbard model (\ref{Eq:BH}) with constant $J_{i,j}=1$ and $N_b=L=7$ for short- ($\alpha=10$) and long-range ($\alpha=0$) couplings. The solid line is the Tracy-Widom distribution and the dashed lines are the best fitting $\chi^{2}$ distributions. The fitting values of $M$ for $\alpha=0$ and $\alpha=10$ are $M=0.94$ and $M=18.0$, respectively (these values are rounded to their closest integers when written in the panels).
}
\label{Fig:5}
\end{figure}

As discussed in the main text, the 1D Bose-Hubbard model (\ref{Eq:BH}) with onsite disorder and a {\em fixed} tunneling parameter shows a ground-state energy distribution that cannot be  reproduced with the expression in Eq.~(\ref{eq: P-Wishart-N1}). The distributions are shown in Fig.~\ref{Fig:5} for short-range ($\alpha=10$) and all-to-all ($\alpha=0$) couplings.

%==============Figure 5 ============
%=================================

We have not yet found an explanation of why the $\chi^2$ distribution is far from $P(E_0)$ for this model, while $P_{\chi^{2}}(E_{0})$ captures  well the shapes for the spin-1/2 model in Eq.~(\ref{Eq:MBL}), which also has a constant coupling strength. This is a point that needs to be further investigated.

\section{Agreement Quantification}
\label{App1}

\begin{figure}[b!]
\centering
\includegraphics[width=1.0\columnwidth]{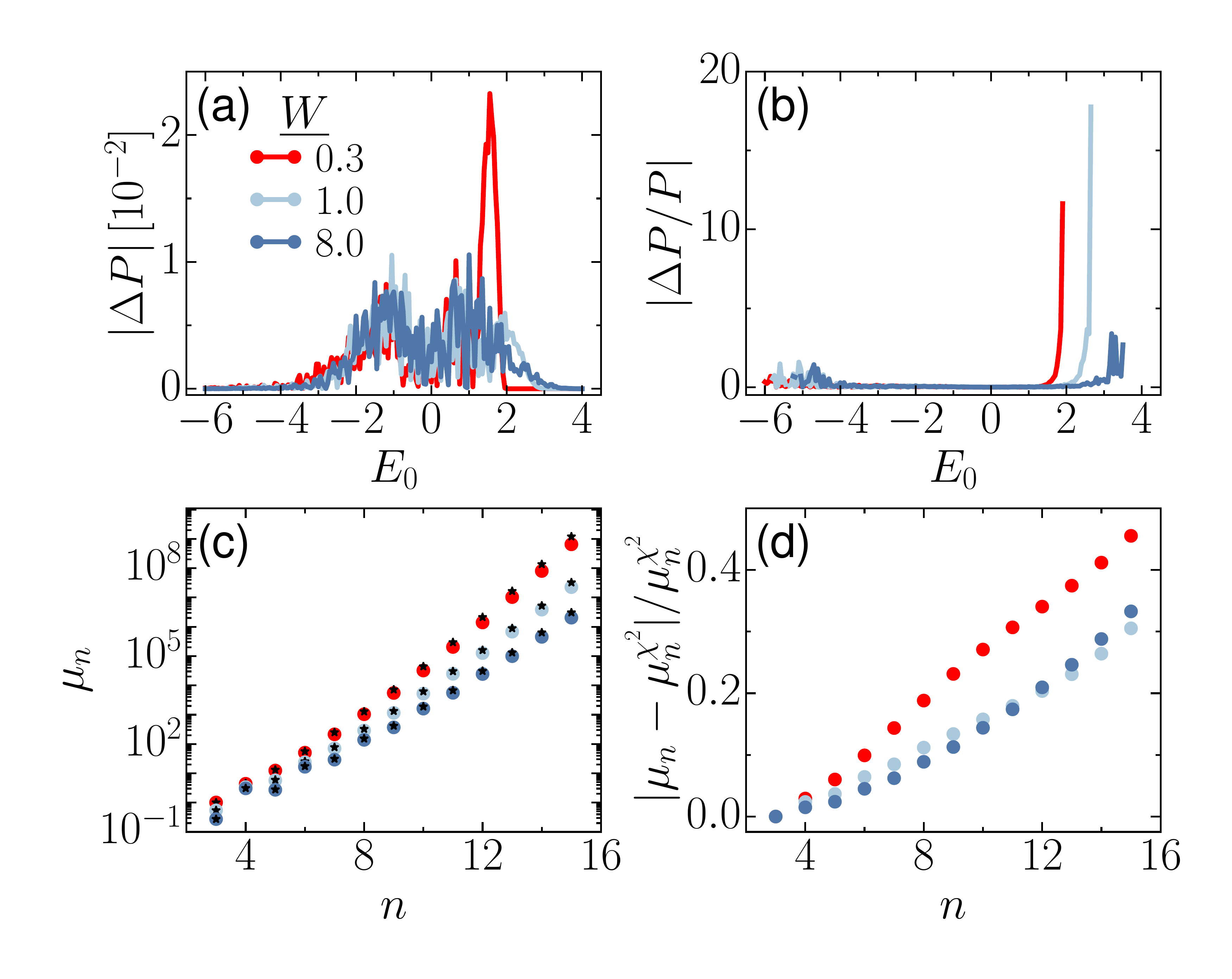}
x\caption{(a) The absolute difference (of the order of $10^{-2}$) between the histograms of the numerical data and their best-fitting $\chi^2$ distributions, $|\Delta P(E_{0})| \equiv |P(E_{0})-P_{\chi^{2}}(E_{0})|$, for the results in panels (b), (d), and (f) of Fig.~\ref{Fig:3}. (b): The absolute relative difference $|\Delta P(E_{0})/P(E_{0})|$ corresponding to the data sets in panel (a). (c): The moments $\mu_n$ as a function of $n$ for $n=3,4,\dots,15$. Moments for the $\chi^2$ distributions are indicated with black stars. (d): Relative difference between the moments of the same data sets in panel (c) and those of their best-fitting $\chi^2$ distributions, $|\mu_n^\text{MBL}-\mu_n^{\chi^2}|/\mu_n^{\chi^2}$, as a function of $n$. In panels (c) and (d), we mirrored all distributions around $E_0 = 0$ such that all moments are positive, and a logarithmic scale can be shown.}
\label{Fig:6}
\end{figure}

In this appendix, we quantify the agreement of the best-fitting $\chi^2$ distribution with the distribution of the ground-state energy shown in Fig.~\ref{Fig:3}. The agreement is overall very good, as quantified by the absolute difference $|\Delta P(E_{0})| \equiv |P(E_{0})-P_{\chi^{2}}(E_{0})|$ between both distributions, with smaller fluctuations at the tails, as shown in Fig.~\ref{Fig:6}(a). This reflects what we could see also with the log plots in Fig.~\ref{Fig:3}. If in Fig.~\ref{Fig:3} we had shown instead linear plots, these small deviations would not have been noticed.

The good agreement between the numerical $P(E_{0})$ and $P_{\chi^{2}}(E_{0})$ is further confirmed by the relative difference shown in Fig.~\ref{Fig:6}(b). Notice that the peak for $E_0>0$ stems from the fact that both $\Delta P(E_{0})$ and $P(E_{0})$ are very close to zero at the positive tail, while the flat part for $E_0<0$ represents the region where $P(E_{0})$ is nonzero. 
For the shifted and scaled distributions, $P_{\chi^{2}}(E_{0})$ vanishes exactly at $E_0=\mu/\sigma=\sqrt{M/2}$, so  the larger deviation close to this point comes as no surprise.

%==============Figure 6 ============
%=================================

In Fig.~\ref{Fig:6}(c) we compare the moments $\mu_n$ of the two distributions. There is very good agreement for lower values of $n$. This is corroborated with the relative difference shown in Fig.~\ref{Fig:6}(d). With increasing $n$, the moments become more sensitive to the tails and an accurate determination of $\mu_n$ requires a very large data set. The large number of samples considered here, $10^6$, may still not be enough for an accurate determination of the highest-$n$ moments. Nevertheless, we observe qualitative agreement also for the large $n$'s.

The extension of the analysis performed in Fig.~\ref{Fig:6} to large and $2\times2$ random matrices provides excellent agreement with the best fitting $\chi^2$-distribution (not shown). For example, the relative differences $|\mu_n - \mu_n^{\chi^2}|/\mu_n^{\chi^2}$ for moments up to $n=15$ are consistently below $10^{-1}$.

%%%%%%%%%%%%%%%%%%%%%%%%%%%%%%%%%%%%%%%%%%%%%%%%

%


\begin{thebibliography}{60}
\bibitem{noteFinch} The value as written in Ref.~[47] is $\gamma = -0.2934645240\dots$.
\bibitem [{foo()}]{footNote}%
  \BibitemOpen
  \href@noop {} {}\bibinfo {note} {For the model in Eq.~(\ref{Eq:MBL}), we also
  observed that for $W\gtrsim 8$ and random numbers from a uniform
  distribution, $P(E_0)$ agrees with Eq.~(\ref{eq:2x2}).}\BibitemShut {Stop}%
\end{thebibliography}
\end{document}